\documentclass[twoside,twocolumn,9pt]{article}
\usepackage{extsizes}
\usepackage[super,sort&compress,comma]{natbib} 
\usepackage[version=3]{mhchem}
\usepackage[left=1.5cm, right=1.5cm, top=1.785cm, bottom=2.0cm]{geometry}
\usepackage{balance}
\usepackage{mathptmx}
\usepackage{sectsty}
\usepackage{graphicx} 
\usepackage{lastpage}
\usepackage[format=plain,justification=justified,singlelinecheck=false,font={stretch=1.125,small,sf},labelfont=bf,labelsep=space]{caption}
\usepackage{float}
\usepackage{fancyhdr}
\usepackage{fnpos}
\usepackage[english]{babel}
\addto{\captionsenglish}{%
  
}
\usepackage{array}
\usepackage{droidsans}
\usepackage{charter}
\usepackage[T1]{fontenc}
\usepackage[usenames,dvipsnames]{xcolor}
\usepackage{setspace}
\usepackage[compact]{titlesec}
\usepackage{hyperref}

\usepackage{ulem}
\usepackage{siunitx}
\newcommand{\Wcm}{\W\per\cm\squared}
\DeclareSIUnit{\au}{a.u.} 

\definecolor{cream}{RGB}{222,217,201}

\begin{document}

\pagestyle{fancy}
\thispagestyle{plain}
\fancypagestyle{plain}{
\renewcommand{\headrulewidth}{0pt}
}

\makeFNbottom
\makeatletter
\renewcommand\LARGE{\@setfontsize\LARGE{15pt}{17}}
\renewcommand\Large{\@setfontsize\Large{12pt}{14}}
\renewcommand\large{\@setfontsize\large{10pt}{12}}
\renewcommand\footnotesize{\@setfontsize\footnotesize{7pt}{10}}
\makeatother

\renewcommand{\thefootnote}{\fnsymbol{footnote}}
\renewcommand\footnoterule{\vspace*{1pt}%
\color{cream}\hrule width 3.5in height 0.4pt \color{black}\vspace*{5pt}} 
\setcounter{secnumdepth}{5}

\makeatletter 
\renewcommand\@biblabel[1]{#1}            
\renewcommand\@makefntext[1]%
{\noindent\makebox[0pt][r]{\@thefnmark\,}#1}
\makeatother 
\renewcommand{\figurename}{\small{Fig.}~}
\sectionfont{\sffamily\Large}
\subsectionfont{\normalsize}
\subsubsectionfont{\bf}
\setstretch{1.125} 
\setlength{\skip\footins}{0.8cm}
\setlength{\footnotesep}{0.25cm}
\setlength{\jot}{10pt}
\titlespacing*{\section}{0pt}{4pt}{4pt}
\titlespacing*{\subsection}{0pt}{15pt}{1pt}

\fancyfoot{}
\fancyfoot[LO,RE]{\vspace{-7.1pt}\includegraphics[height=9pt]{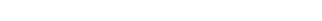}}
\fancyfoot[CO]{\vspace{-7.1pt}\hspace{11.9cm}\includegraphics{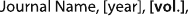}}
\fancyfoot[CE]{\vspace{-7.2pt}\hspace{-13.2cm}\includegraphics{head_foot/RF}}
\fancyfoot[RO]{\footnotesize{\sffamily{1--\pageref{LastPage} ~\textbar  \hspace{2pt}\thepage}}}
\fancyfoot[LE]{\footnotesize{\sffamily{\thepage~\textbar\hspace{4.65cm} 1--\pageref{LastPage}}}}
\fancyhead{}
\renewcommand{\headrulewidth}{0pt} 
\renewcommand{\footrulewidth}{0pt}
\setlength{\arrayrulewidth}{1pt}
\setlength{\columnsep}{6.5mm}
\setlength\bibsep{1pt}

\makeatletter 
\newlength{\figrulesep} 
\setlength{\figrulesep}{0.5\textfloatsep} 

\newcommand{\topfigrule}{\vspace*{-1pt}%
\noindent{\color{cream}\rule[-\figrulesep]{\columnwidth}{1.5pt}} }

\newcommand{\botfigrule}{\vspace*{-2pt}%
\noindent{\color{cream}\rule[\figrulesep]{\columnwidth}{1.5pt}} }

\newcommand{\dblfigrule}{\vspace*{-1pt}%
\noindent{\color{cream}\rule[-\figrulesep]{\textwidth}{1.5pt}} }

\makeatother

\twocolumn[
  \begin{@twocolumnfalse}
\vspace{1em}
\sffamily
\begin{tabular}{m{4.5cm} p{13.5cm} }

 & \noindent\LARGE{\textbf{Electron recollisional excitation of OCS$^+$ in phase-locked $\omega + 2\omega$ intense laser fields}} \\
\vspace{0.3cm} & \vspace{0.3cm} \\

 & \noindent\large{Tomoyuki Endo,$^{\ast}$\textit{$^{a}$} Tomohito Otobe,\textit{$^{a}$} and Ryuji Itakura\textit{$^{a}$}} \\

& \noindent\normalsize{Photoelectron-photoion coincidence momentum imaging has been performed to investigate excitation processes on dissociative ionization of OCS, OCS $\to$ OCS$^+$ + e$^-$ $\to$ OC + S$^+$ + e$^-$, in phase-locked $\omega + 2\omega$ intense laser fields.
The electron kinetic energy spectra depend on the ion species, OCS$^+$ or S$^+$, produced in coincidence.
The observed electron momentum distribution shows clear asymmetry along the laser polarization direction with a 2$\pi$-oscillation period as a function of the phase difference between the $\omega$ and $2\omega$ laser fields.
The asymmetry of electron emission in the OCS$^+$ channel flips at the electron kinetic energy of \SI{8.2}{\eV}, below which forward-scattered electrons dominate and above which backward-scattered electrons dominate.
In the S$^+$ channel, the asymmetry flips at the lower kinetic energy of \SI{4.2}{\eV}.
In comparison with a classical trajectory Monte Carlo simulation, it has been clarified that this energy shift between the OCS$^+$ and S$^+$ channels corresponds to the excitation energy of the parent ion and that electron recollisional excitation takes place to form the fragment ion in intense laser fields.} \\

\end{tabular}

 \end{@twocolumnfalse} \vspace{0.6cm}

  ]

\renewcommand*\rmdefault{bch}\normalfont\upshape
\rmfamily
\section*{}
\vspace{-1cm}


\footnotetext{\textit{$^{a}$~Kansai Institute for Photon Science, National Institutes for Quantum Science and Technology, 8-1-7 Umemidai, Kizugawa, Kyoto 619-0215, Japan. E-mail: endo.tomoyuki@qst.go.jp}}




\section{Introduction}

Tunneling ionization and subsequent electron recollision processes in intense laser fields induce ultrafast molecular and electronic dynamics such as high-order harmonic generation~\cite{Itatani.Nature.2004, Baker.Science.2006, Worner.Nature.2010}, elastic electron scattering~\cite{Meckel.Science.2008, Blaga.Nature.2012, Pullen.NatComm.2015, Giovannini.JPB.2023}, electronic excitation leading to molecular dissociation~\cite{Niikura.Nature.2003, Kling2006, Li.JPhysB.2014, Endo.PRL.2016, Wanie2016, Alnaser2014}, and non-sequential double ionization~\cite{Hishikawa.PRL.2007, Bergues.NatCommun.2012}.
The tunneling electron gains kinetic energy in alternating laser electric fields, and the excess energy causes such non-linear phenomena.
The tunneling ionization and electron recollision processes occur within a single optical cycle (\SI{2.7}{\fs} at \SI{800}{\nm}) as described in a three-step model~\cite{Corkum.PRL.1993, Lewenstein.PRA.1994}, thus these subsequent phenomena are promising for imaging ultrafast molecular and electronic dynamics~\cite{Meckel.Science.2008, Kraus.ChemPhys.2013, Xu.JPB.2016}.
Since the tunneling electron selectively collides with the parent ion, the electron recollision in intense laser fields is a powerful tool to investigate elastic/inelastic electron scattering by ions.

One of the applications of laser-induced electron scattering processes is laser-induced electron diffraction (LIED), which is a promising candidate to investigate ultrafast nuclear dynamics with attosecond and picometer resolutions with an elastic scattering process~\cite{Meckel.Science.2008, Blaga.Nature.2012, Pullen.NatComm.2015, Giovannini.JPB.2023}.
The structure and dynamics of molecules at the time when the electron recollides are encoded in the rescattered electron wavepacket.
The contribution of an electron recollision process in LIED has been verified by measuring the maximum kinetic energy of the backward-scattered electron~\cite{Trabattoni.NatComm.2020}.

Another application of laser-induced electron scattering processes is electron recollisional ionization and excitation.
The contributions of inelastic scattering in double ionization in intense laser fields have been discussed from the correlation between electron momenta of the first and second electrons~\cite{Bergues.NatCommun.2012}.
The double ionization mechanisms such as non-sequential double ionization (NSDI) or recollision-induced excitation and subsequent field ionization (RESI) have been proposed~\cite{Feuerstein.PRL.2001, Bergues.IEEE.2015}.
The contribution of an electron recollision process to form electronically excited cationic states leading to molecular dissociation has been discussed on the basis of dependencies on ellipticity~\cite{Niikura.Nature.2002, Endo.PRL.2016} and carrier-envelope-phase (CEP)~\cite{Xie.PRL.2012} of laser fields.
However, it remains challenging to identify which electronically excited cationic states are populated by electron recollision.

The electron-nuclear joint energy spectrum has been proposed to reveal the energy correlation between the photoelectron and nuclear ions~\cite{Wu.PRL.2013, Sun.PRA.2016, Lu.PNAS.2018}.
However, it is difficult to apply to polyatomic molecules due to the complex nuclear dynamics, which lead to the redistribution of internal energy into multiple vibrational and rotational modes.

Inelastic scattering is difficult to elucidate solely from electron kinetic energy partly because of poor statistics in the cutoff region, where too few electrons are detected to reliably resolve small differences in cutoff energies associated with different electronic states.
In addition, other processes such as tunneling ionization from lower lying orbitals would compete.
The total kinetic energy release of fragments, which reflects the difference between the potential energies of the excited state at the excitation and at the dissociation limit, is one of the useful observables to discuss which electronic states involved in molecular dissociation in intense laser fields.
Other observables such as three-dimensional momentum distributions including angular distributions, intensity dependence, and wavelength dependence also provide insight into the dissociation pathways, but it is not easy to assign the electronic states populated in intense laser fields because dissociation pathways in intense laser fields are complicated by the light-induced deformation of potential energy surfaces~\cite{Kling2006, Sato2003, Kono.BCSJ.2006} and light-matter interaction after ionization~\cite{Endo.FrontInChem.2022}.

\begin{figure}[ht!]
    \centering
    \includegraphics[width=\hsize]{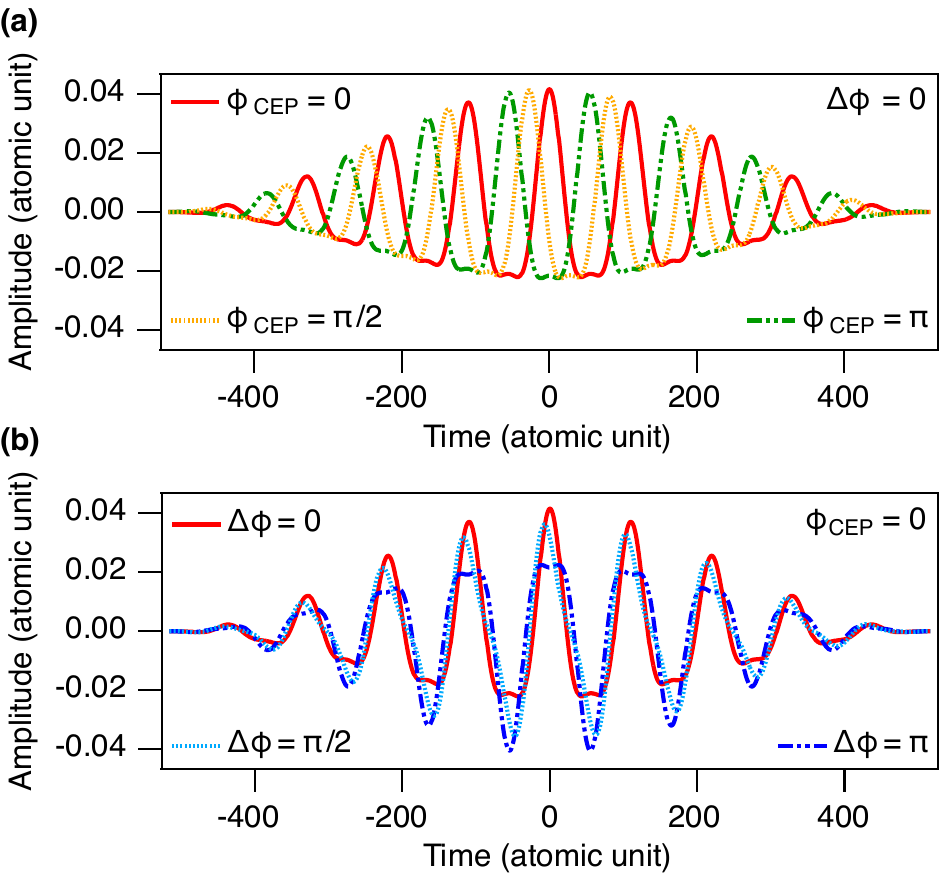}
    \caption{
    Examples of phase-locked $\omega + 2\omega$ laser electric fields for different (a) $\phi_\text{CEP}$ and (b) $\Delta \phi$.
    }
    \label{fig:laserfields}
\end{figure}

Using phase-locked two-color ($\omega + 2\omega$) laser fields is a simple and powerful method to obtain spatial asymmetry in laser fields and to investigate the ionization and dissociation mechanisms irrespective of polar or non-polar molecules~\cite{Song2015, Doblhoff-Dier2016, Endo.JESRP.2016, Endo.PCCP.2017, Hasegawa2022}.
An electric field of a phase-locked $\omega + 2\omega$ laser field can be expressed as
\begin{eqnarray}
    F(t, \Delta\phi) & = & F_\omega (t) \cos \left( \omega t + \phi_\text{CEP} \right)  \nonumber \\
    & + & F_{2\omega} (t) \cos \left( 2 \omega t + 2\phi_\text{CEP} + \Delta \phi \right) , \label{eq:TwoColorField}
\end{eqnarray}
where $F_\omega (t)$ and $F_{2\omega} (t)$ are the envelope functions of laser fields with the carrier frequencies of $\omega$ and $2\omega$, respectively, $\phi_\text{CEP}$ is the CEP of the fundamental $(\omega)$ pulse, and $\Delta \phi$ is the phase difference between the $\omega$ and second harmonic $(2\omega)$ pulses.
Examples of the $\omega + 2\omega$ laser electric fields at different combinations of $\phi_\text{CEP}$ and $\Delta \phi$ are shown in Figs.~\ref{fig:laserfields}(a) and (b), respectively.
The asymmetry of the laser electric fields depends on $\Delta \phi$, but not on $\phi_\text{CEP}$.
When the multi-cycle one-color laser fields are employed, the electrons are ejected symmetrically along the laser polarization direction.
In contrast, the phase-locked $\omega + 2\omega$ laser fields enable us to obtain the asymmetric electron and fragment momentum distributions along the laser polarization direction~\cite{Song2015, Doblhoff-Dier2016, Endo.JESRP.2016, Endo.PCCP.2017, Hasegawa2022}.

In this paper, photoionization of carbonyl sulfide (OCS) molecules in the phase-locked $\omega + 2\omega$ intense laser fields has been investigated by the photoelectron-photoion coincidence (PEPICO) three-dimensional momentum imaging technique to clarify the role of electron recollision in populating electronically excited cationic states in intense laser fields.
The direction of electron ejection and its dependence on the shapes of laser electric fields have been measured and compared with a classical trajectory Monte Carlo (CTMC) simulation with a potential energy surface for a photoelectron calculated using density functional theory (DFT).

\begin{figure}[ht!]
    \centering
    \includegraphics[width=\hsize]{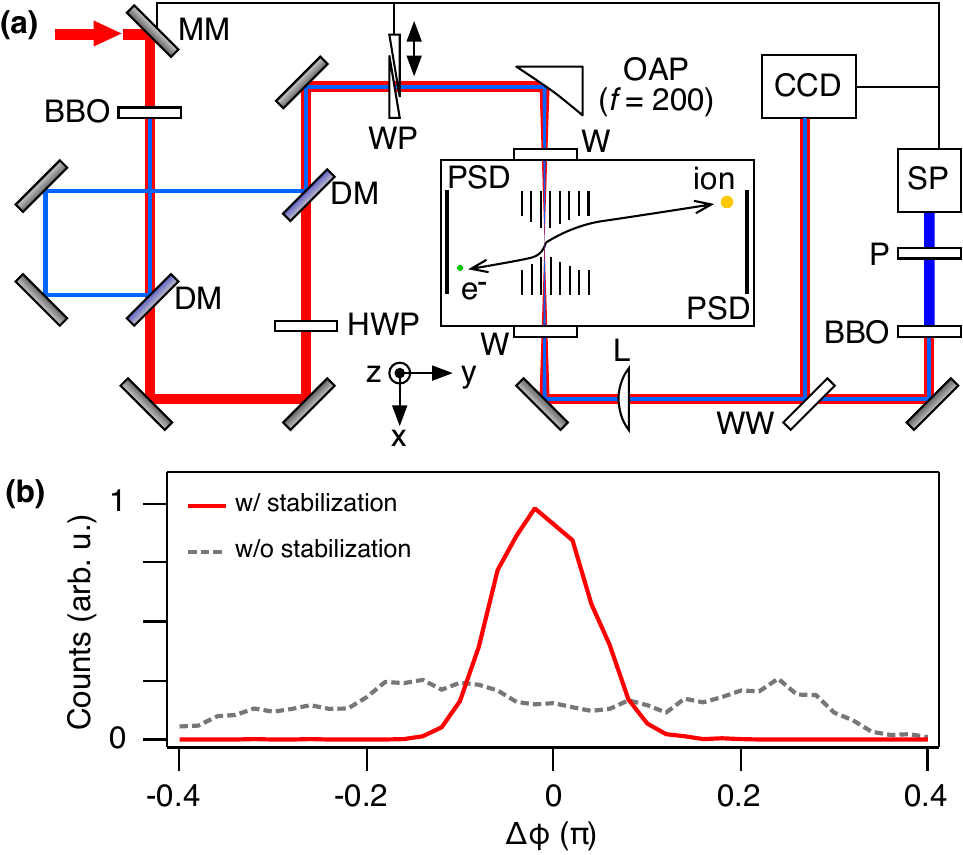}
    \caption{
    (a) Schematic of the experimental setup. MM: motorized mount; BBO: $\beta$-barium borate crystal; DM: dielectric mirror; HWP: half-waveplate; WP: wedge plate; OAP: off-axis parabolic mirror; W: window; PSD: position-sensitive detector; L: lens; WW: wedged window; CCD: charge-coupled device camera; P: polarizer; SP: spectrometer.
    (b) Distributions of phase difference $\Delta \phi$ measured with feedback-loop stabilization (red solid line) and without stabilization (grey broken line).
    }
    \label{fig:experimental}
\end{figure}

\section{Experiment}
A schematic diagram of the experimental setup is shown in Fig.~\ref{fig:experimental}(a).
The output of a Ti:Sapphire chirped pulse amplifier ($\lambda \sim$ $\SI{800}{\nm}$, \SI{70}{\fs}, \SI{1}{\kilo\Hz}) is introduced into a $\beta$-barium borate (BBO, type I, cut angle $\theta = \SI{29.2}{\deg}$, \SI{0.1}{\mm} thickness) crystal to generate the $2\omega$ pulses ($\lambda \sim$ \SI{400}{\nm}).
The $\omega$ and $2\omega$ pulses are separated by a dielectric mirror (Layertec, 106853), which reflects the $2\omega$ pulses and transmits the $\omega$ pulses.
The polarization direction of the $\omega$ pulses is rotated by a half-waveplate to be parallel to that of the $2\omega$ pulses.
The $\omega$ and $2\omega$ pulses are co-linearly combined again by another dielectric mirror and focused by an off-axis parabolic mirror ($f = \SI{200}{\mm}$) onto an effusive molecular beam of a gas mixture (He 95\% + OCS 5\%) introduced into an ultrahigh vacuum chamber (residual gas pressure $< \SI{e-8}{\Pa}$).
The spatial overlap of the $\omega$ and $2\omega$ pulses are optimized to maximize the number of the electron yields.
The effective laser field intensity of the $\omega$ and $2\omega$ pulses at the focal spot is estimated to be \SI{5e13}{\Wcm} and \SI{5e12}{\Wcm}, respectively, from the ponderomotive energy shift $U_p$ measured in the photoelectron spectra of Xe.
The phase difference $\Delta\phi$ at the focal spot is calibrated by measuring the phase difference dependence of the maximum kinetic energy of backward-scattered electrons from Xe~\cite{Ray.PRA.2011}.

To stabilize $\Delta \phi$ over a long acquisition period, a feedback system is used in the present study.
The $\omega + 2\omega$ laser pulses passed through the experimental chamber are introduced into another $\beta$-BBO crystal to generate the $2\omega$ pulses from the residual $\omega$ pulses.
The shift of $\Delta \phi$ is monitored by measuring the interference spectrum between the $2\omega$ pulses generated at two $\beta$-BBO crystals, being compensated by using a pair of fused silica wedge plates mounted on a linear motorized stage~\cite{Endo.JESRP.2016, Endo.PCCP.2017, Endo.PRA.2019, Endo.FrontInChem.2022}.
Distributions of $\Delta \phi$ measured for 15 minutes with and without stabilization are shown in Fig.~\ref{fig:experimental}(b).
The measured $\Delta \phi$ without stabilization is distributed over 0.5$\pi$, which is large enough to change the shape of the laser electric fields from asymmetric to symmetric (see Fig.~\ref{fig:laserfields}(b)).
The standard deviation of $\Delta \phi$ with stabilization is kept below 0.06$\pi$ over a week.
A part of the $\omega + 2\omega$ laser pulses is reflected by a wedged window to monitor and stabilize the beam position at the focal spot by using a charge-coupled device camera and a motorized mirror mount.

Details of the PEPICO momentum imaging system used in this study are described elsewhere~\cite{Hosaka.JCP.2013, Ikuta.PRA.2022, Fukahori.PRA.2023}.
Briefly, the generated ions and electrons in the phase-locked $\omega + 2\omega$ laser fields are accelerated to the opposite directions by an electrostatic field, and detected by using respective fast micro-channel plate detectors with position-sensitive delay-line anodes (RoentDek HEX80).
The three-dimensional momentum vector of each charged particle $\mathbf{p} = (p_x, p_y, p_z)$ is obtained by measuring the position on the detector ($x, z$) and the arrival time $t$ in a single-shot acquisition mode.
The event rate is kept at about 0.3 per shot to reduce false events.

\section{Simulation}
The electron momentum distribution in the phase-locked $\omega + 2\omega$ laser fields is simulated with a CTMC method~\cite{Wolter.PRA.2014, Hao.PRA.2020}.
The following equations are given in atomic units.
Here, we employ a $\sin^2$ envelope for both pulses for simplicity.
Thus, the envelope functions in Eq.~(\ref{eq:TwoColorField}) are given by
\begin{eqnarray}
    F_\omega (t) = F_\omega \sin^2 \left( \pi t / \tau_\omega \right), \\
    F_{2\omega} (t) = F_{2\omega} \sin^2 \left( \pi t / \tau_{2\omega} \right),
\end{eqnarray}
where $F_\omega$ and $F_{2\omega}$ are the amplitudes of the laser electric fields, and $\tau_\omega$ and $\tau_{2\omega}$ are the full width of the $\omega$ and $2\omega$ pulses.
The durations of $\tau_\omega = \tau_{2\omega} = \SI{1033.54}{\au}$ (\SI{25}{\fs}) are employed in the simulation to reduce the computational cost.
Since ionization occurs mainly near the peak of the envelope, the CTMC results are insensitive to the pulse duration in the multi-cycle regime.

The initial electron flux is obtained as the tunneling ionization rate from the highest occupied molecular orbital (HOMO) based on weak-field asymptotic theory~\cite{Tolstikhin.PRA.2011, Madsen.PRA.2012}, which includes the effects of permanent dipole moment as well as the shape of HOMO.
The tunneling ionization rate $\Gamma(\beta, F)$ as a function of the molecular orientation angle $\beta$ with respect to the $z$-axis, which is parallel to the laser polarization direction, and the fields strength $F$ can be expressed as
\begin{eqnarray}
    \Gamma(\beta, F) & = & \left[ |G_{00}(\beta)|^2 + \frac{F}{2\kappa^2} |G_{01}(\beta)|^2 \right] W_{00}(F), \label{eq:WFAT} \\
    W_{00}(F) & = & \frac{\kappa}{2} \left( \frac{4 \kappa^2}{F} \right)^{2/\kappa - 1} \exp \left( - \frac{2\kappa^3}{F} \right),
\end{eqnarray}
where $G_{00}(\beta)$ and $G_{01}(\beta)$ are the structure factors describing the dependence of ionization rate on the molecular orientation with respect to the laser electric fields due to the shape of a molecular orbital and a dipole momentum, $W_{00}(F)$ is the field factor, and $\kappa = \sqrt{2I_p}$ with $I_p$ being the ionization potential of OCS.
The structure factors of OCS are taken from~\cite{Madsen.PRA.2013}.
The molecular orientation $\beta$ is chosen randomly from a constant $\cos\beta$ distribution over the [-1, 1] range in each trajectory simulation.
By taking into account the effect of depletion of neutral molecules, the effective tunneling ionization rate $\Gamma_{\rm eff}$ at the ionizing time $t_i$ with the time dependent laser electric fields $F(t)$ can be expressed as
\begin{equation}
    \Gamma_{\rm eff} (\beta, F(t_i)) = \left[ 1 - \int_{-\infty}^{t_i} \Gamma_{\rm eff} (\beta, F(t)) dt \right] \Gamma(\beta, F(t_i)).
\end{equation}

\begin{figure}
    \centering
    \includegraphics[width=\hsize]{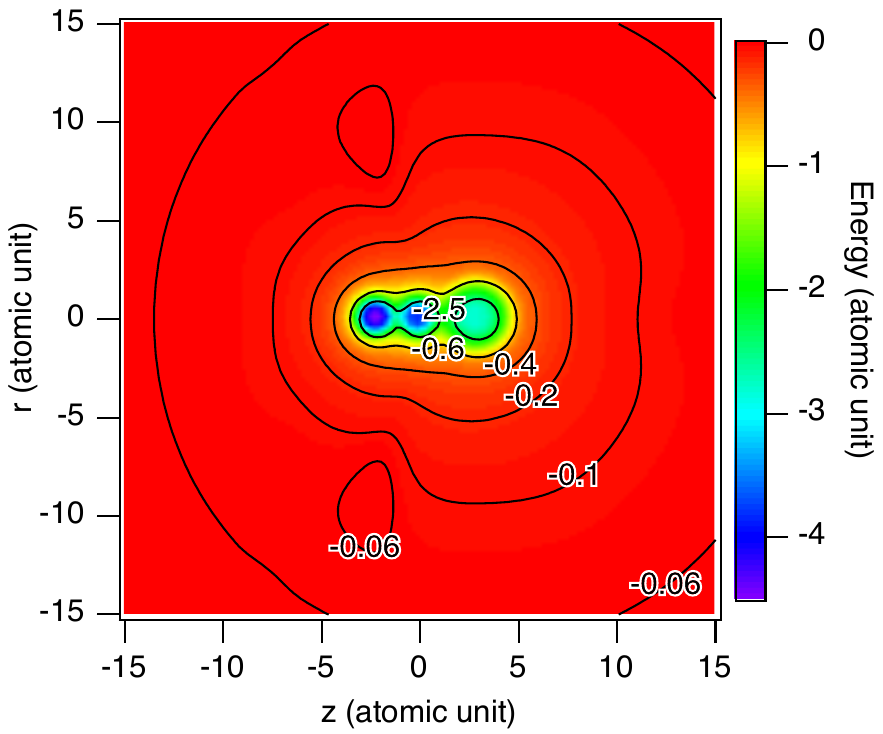}
    \caption{
    Two-dimensional plot of the potential energy $V(\beta = 0, r, z)$ employed in the CTMC simulation at $\beta = 0$.
    The position of each nucleus is at the equilibrium position in neutral OCS.
    The carbon atom is at the origin, the oxygen atom is at $z < 0$, and the sulfur atom is at $z > 0$.
    The laser polarization direction is along the $z$-axis.}
    \label{fig:Potential}
\end{figure}

The position and velocity of the tunneling electron as a function of the time are calculated by sequentially solving the classical equations of motion:
\begin{eqnarray}
    \frac{d^2 r}{dt^2} & = & -\frac{d}{dr} V(\beta, r, z), \label{eq:eq_motion_r}\\ 
    \frac{d^2 z}{dt^2} & = & F(t, \Delta \phi) - \frac{d}{dz} V(\beta, r, z), \label{eq:eq_motion_z} 
\end{eqnarray}
where $r, z$ are the perpendicular and parallel components of cylindrical coordinates with respect to the laser polarization direction, respectively, $V(\beta, r, z)$ is a potential energy of the electronic ground state of OCS$^+$, and $F(t, \Delta \phi)$ is the $\omega + 2\omega$ laser electric fields given by Eq.~(\ref{eq:TwoColorField}).

The potential energy surface $V(\beta, r, z)$ is obtained by combining a DFT potential $V_\text{DFT}$ for the inner region ($|r| < \SI{5.9}{}, \SI{-5.9}{} < z < \SI{10.6}{}$) and a three-center Coulomb potential $V_\text{Coulomb}$ for the outer region ($|r| > \SI{14}{}, z < \SI{-9}{}, \SI{18}{} < z$).
The middle region is interpolated with a cubic spline.
The DFT potential is calculated with the Krieger-Li-Iafrate self-interaction correlation~\cite{Tong.PRA.1997}.
The three-center Coulomb potential $V_\text{Coulomb}$ is given by
\begin{eqnarray}
    V_\text{Coulomb} = - \sum_{i} \frac{q_i}{\sqrt{(r - r_i)^2 + (z - z_i)^2}},
\end{eqnarray}
where $q_i$, $r_i$, and $z_i$ are a charge and positions of the $i$-th atom, respectively.
Net charge on each atom is evaluated as the electron distribution perceived by a HOMO electron: $-0.30$, 0.55, and 0.75 for the oxygen, carbon, and sulfur atoms, respectively.
The employed potential for $\beta = 0$ is shown in Fig.~\ref{fig:Potential}.
The position of each nucleus is at the equilibrium position in the neutral OCS and fixed in the simulation because the tunneling ionization and rescattering processes occur within an optical-cycle (\SI{2.7}{\fs}), which is shorter than a vibrational period of a C-S stretching mode (\SI{874}{\cm^{-1}}, \SI{38}{\fs}).
The carbon atom is at the origin, the oxygen atom is at $z < 0$, and the sulfur atom is at $z > 0$.

The initial position of the electron at the ionizing time $t_i$ along the $z$-axis is determined as the tunnel exit point given by solving the equation $zF(t_i, \Delta\phi) + V(\beta, r = 0, z) = -I_p$.
The distribution of the initial transverse velocity~\cite{Pham.PRA.2014} is given as 
\begin{equation}
    w(v_r, t_i) = \frac{4\pi \kappa}{F(t_i, \Delta\phi)} \exp \left[- \frac{\kappa v_r(t_i)^2}{F(t_i, \Delta\phi)} \right]. \label{eq:transverse_vel}
\end{equation}
The other initial conditions are $r(t_i) = 0$, $v_z(t_i) = 0$.
In this simulation, $t_i$, $\phi_\text{CEP}$, $\beta$, and $v_r$ are randomly chosen from constant distributions in the range of $t_i=[0, 1033.54], \phi_\text{CEP}=[-\pi, \pi], \cos\beta=[-1, 1],$ and $v_r = [-4\sqrt{(F_\omega + F_{2\omega})/\kappa}, 4\sqrt{(F_\omega + F_{2\omega})/\kappa}]$ for each trajectory.
According to Eq.~(\ref{eq:transverse_vel}), the probability to take $v_r = \pm 4\sqrt{(F_\omega + F_{2\omega})/\kappa}$ is $\exp(-16)$ at maximum.

In order to take an inelastic scattering process into account, a boundary is determined as the position where the potential energy is equal to the field-free ionization potential of OCS$^+$, $V_\text{b}(\beta, r, z) = - I_{p,2} = \SI{-1.11}{\au} (\SI{-30.3}{\eV})$, and the energy transfer between the photoelectron and the parent ion is assumed to occur at an incoming boundary.
When the electron, which has larger kinetic energy than the threshold energy for the $A$-$X$ transition of OCS$^+$, approaches the parent ion across the boundary, the electron loses the excitation energy of $E_\text{th} = \SI{4}{\eV}$~\cite{Orth.CP.1980}.
The relationship between the kinetic energies before and after the inelastic scattering can be written as
\begin{eqnarray}
    \frac{1}{2} v_r^2(t) + \frac{1}{2} v_z^2(t) & = & \frac{1}{2} v_r'^2(t) + \frac{1}{2} v_z'^2(t) + E_\text{th},
\end{eqnarray}
where $v_r'(t)$ and $v_z'(t)$ are the velocity along the $r$ and $z$ axes, respectively, after the inelastic scattering.
To evaluate the effects of angular redistribution of electron momenta caused by the inelastic scattering, two models were employed:
One is a model without angular redistribution.
This model assumes that the propagation direction of an inelastically scattered electron remains unchanged at the moment of the energy transfer.
Any subsequent change in propagation direction, compared with the elastic scattering process, arises from the Coulomb interaction with a decelerated electron.
Another is a model with random angular redistribution by inelastic scattering.
This model assumes that the propagation direction of an electron is redistributed isotropically over a full 2$\pi$ range due to inelastic scattering.

The propagation is performed using a standard fourth-order Runge-Kutta algorithm and proceeds until the time of \SI{16e5}{\au} sufficiently after the end of the laser pulse (\SI{1033.54}{\au})~\cite{Wolter.PRA.2014}.
After the propagation, the momentum of the electron with a distance from the origin larger than \SI{10000}{\au} is used for the following analysis.

\begin{figure}
    \centering
    \includegraphics[width=\hsize]{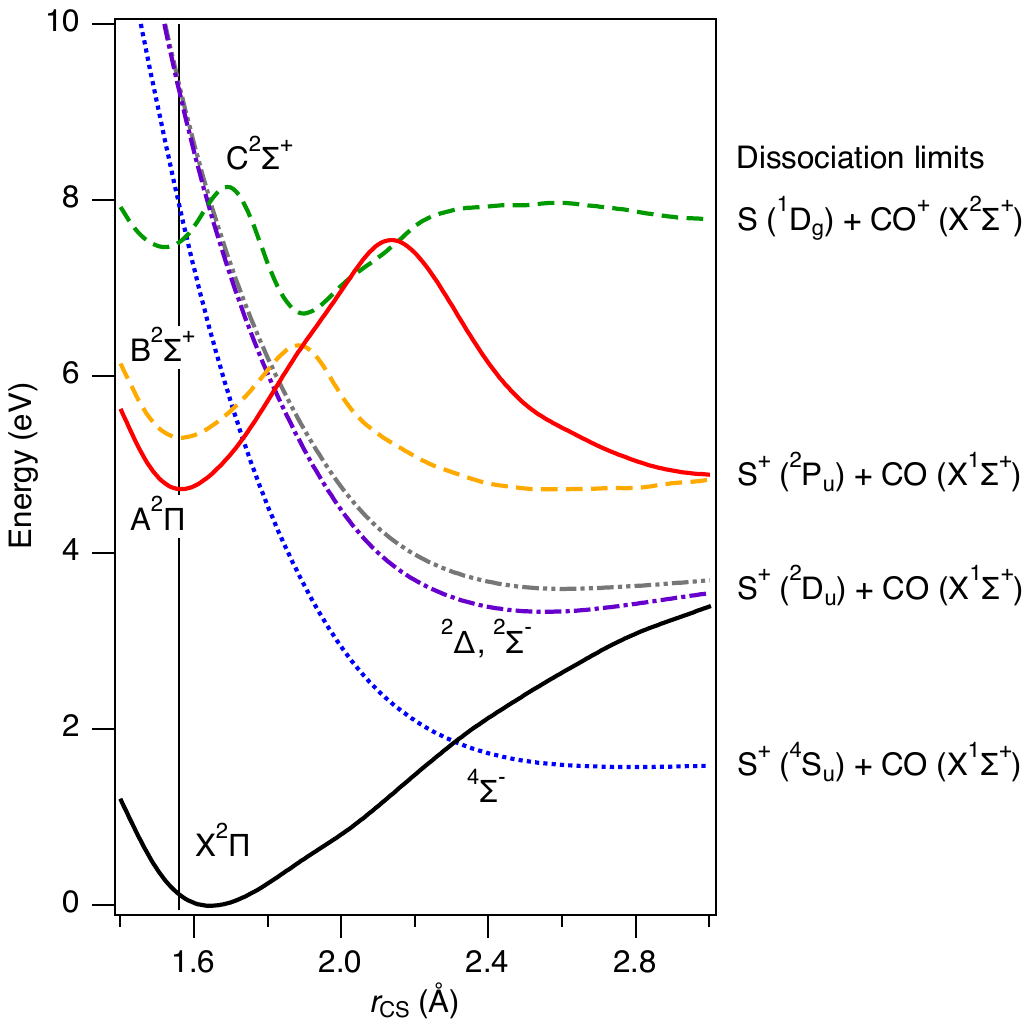}
    \caption{
    Schematic of potential energy curves of OCS$^+$ along the C--S coordinate ($r_\text{CO} = \SI{1.15}{\angstrom}$) adapted from Ref.~\citenum{Hirst.MolPhys.2006} by digitizing the published data.
    Solid, dashed, dotted, dash-dotted, and dash-dot-dotted lines represent $^2\Pi$, $^2\Sigma^+$, $^4\Sigma^-$, $^2\Delta$, and $^2\Sigma^-$ states, respectively.
    Vertical line indicates the equilibrium geometry of OCS.
    }
    \label{fig:PotentialCurves}
\end{figure}

\section{Results and Discussion}
Here, we focus on the following two major ionization channels;
(i) OCS $\to$ OCS$^+$ + e$^-$,
(ii) OCS $\to$ OCS$^{+*}$ + e$^-$ $\to$ OC + S$^+$ + e$^-$.
The former is the OCS$^+$ channel, and the latter is the S$^+$ channel.
Since the ground $X^2\Pi$ state of OCS$^+$ is a stable state, S$^+$ would be generated via electronically excited states OCS$^{+*}$ such as $A^2\Pi$ or $B^2\Sigma^+$ states, which are lying about \SI{4}{}-\SI{5}{\eV} above the $X^2\Pi$ state of OCS$^+$~\cite{Orth.CP.1980}.
Potential energy curves for relevant electronic states along the C--S bond coordinate~\cite{Hirst.MolPhys.2006} are shown in Fig.~\ref{fig:PotentialCurves}.
Previous studies proposed that the dissociation of OCS$^+$ via the $A^2\Pi$ state occurred either directly to the repulsive states or through internal conversion to the vibrationally hot $X^2\Pi$ state~\cite{Wang.JPCA.2024}, and via the $B^2\Sigma^+$ state occurred through internal conversion to the $X^2\Pi$ and $A^2\Pi$ states~\cite{Chang.JPCA.2005}.
Since the kinetic energies of the fragments via the $A^2\Pi$ and $B^2\Sigma^+$ states are less than 1 eV~\cite{Chang.JPCA.2005, Wang.JPCA.2024} and broadly distributed by intramolecular vibrational redistribution, it is difficult to distinguish them from the fragment kinetic energy in the present study.
Since no faster S$^+$ fragments with kinetic energies exceeding \SI{1}{\eV} are observed, contributions from higher-lying electronic states such the $C^2\Sigma^+$ state would be minor.
In our experimental conditions, the yield of the S$^+$ channel is about 5\% of that of the OCS$^+$ channel.
Other channels, e.g. OC$^+$ and CS$^+$ channels, are minor.

\subsection{Measurement of channel-resolved photoelectron momenta}

\begin{figure}
    \centering
    \includegraphics[width=\hsize]{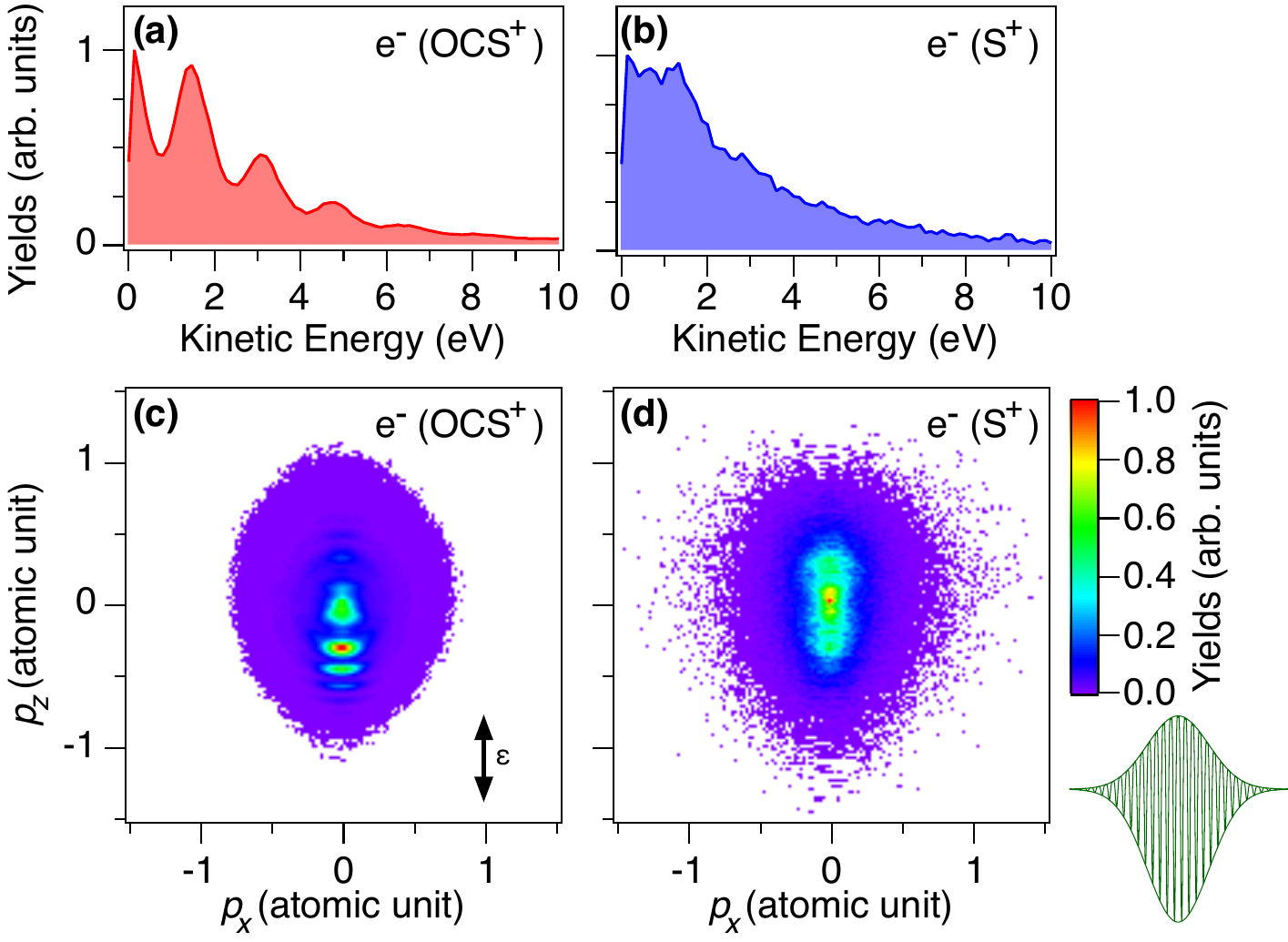}
    \caption{Electron kinetic energy $E_\text{kin}$ distribution of (a) the OCS$^+$ channel and (b) the S$^+$ channel obtained by averaging over $\Delta \phi$ from 0 to 2$\pi$. 
    Momentum images of the photoelectron coincidentally detected with (c) OCS$^+$ and (d) S$^+$ in phase-locked $\omega + 2\omega$ laser fields ($I_\omega = \SI{5e13}{\Wcm}, I_{2\omega} = \SI{5e12}{\Wcm}$) at $\Delta \phi = \pi$.
    The laser polarization direction is indicated as $\varepsilon$.
    The shape of the $\omega + 2\omega$ laser electric field at $\Delta \phi = \pi$ is schematically shown at the bottom-right corner.}
    \label{fig:momimage}
\end{figure}

The electron kinetic energy $E_\text{kin}$ distributions are obtained by averaging over $\Delta \phi$ from 0 to $2\pi$.
The phase-averaged $E_\text{kin}$ distribution of the OCS$^+$ channel in Fig.~\ref{fig:momimage}(a) has a series of peaks with the interval of \SI{1.55}{\eV}, which corresponds to the photon energy of the $\omega$ pulses.
The periodic peak structure can be attributed to above threshold ionization~\cite{Agostini.PRL.1979}, which can be described not only by the multiphoton picture but also by periodic tunneling ionization in time.
On the other hand, the $E_\text{kin}$ distribution of the S$^+$ channel in Fig.~\ref{fig:momimage}(b) has only a broad peak.

The following mechanisms for the S$^+$ channel are possible as combinations of ionization and excitation processes;
\begin{enumerate}
    \item[I)] multi-photon or tunneling ionization to the ground state of OCS$^+$, and multi-photon excitation to the excited state,
    \item[II)] tunneling ionization to the excited state (tunneling ionization from the inner valence orbital, e.g. HOMO-1),
    \item[III)] tunneling ionization to the ground state, and electron recollisional excitation to the excited state.
\end{enumerate}
If the mechanism I is dominant, since the photoelectrons are not affected by the excitation process after photoelectron ejection, the periodic peak structure should also appear in the $E_\text{kin}$ spectrum of the S$^+$ channel as well as the OCS$^+$ channel.
The disappearance of the periodic peak structure means that the different ionization process from the OCS$^+$ channel or coupling between the ionization and excitation processes play an important role in the S$^+$ channel.

The Keldysh parameter in the present experimental condition can be estimated to be 1.38 from the measured parameters, which is attributed to the intermediate region between the multiphoton and tunneling ionization.
For polyatomic molecules, multiphoton ionization and tunneling ionization have been known to coexist when the Keldysh parameter is moderately larger than 1~\cite{Levis.JPCA.1999}.
In fact, in the case of CS$_2$, the multiphoton ionization contributes to the formation of the parent ions and the rescattering process contributes to the production of excited states or doubly charged ions~\cite{Matsuda.RSI.2011}.

In the experiment for OCS~\cite{Shi.ChinPhysB.2024}, a knee structure is observed around 4-5$\times 10^{13}$ W/cm$^2$ in the intensity dependence of the S$^+$ fragment yields, indicating that the non-sequential process plays an important role in producing the S$^+$ fragments at such intensities.
The ellipticity dependence also observed at the intensity of \SI{1.5e14}{\Wcm}~\cite{Shi.ChinPhysB.2024} suggests that recollision of a tunneling electron may play an important role in the fragmentation in near-infrared laser fields.
To clarify the mechanism further, the dependence on the shape of the laser electric fields will be investigated.

Figures~\ref{fig:momimage}(c) and (d) show the sliced images of the three-dimensional momenta ($|p_y| < 0.05$ atomic units) of the photoelectrons of the OCS$^+$ and S$^+$ channels, respectively, in the phase-locked $\omega + 2\omega$ intense laser fields ($I_\omega = \SI{5e13}{\Wcm}, I_{2\omega} = \SI{5e12}{\Wcm}$) at $\Delta\phi = \pi$.
The shape of the laser electric field at $\Delta \phi = \pi$ is schematically shown at the bottom right corner of Fig.~\ref{fig:momimage}.
The electron momentum images show anisotropic distributions with respect to the laser polarization direction $\varepsilon$ along the $z$-axis.
The electron image of the OCS$^+$ channel shows a periodic peak structure corresponding to a series of peaks appearing in the $E_\text{kin}$ distribution and an asymmetric distribution with larger electron yields on the larger amplitude side of the laser electric fields ($p_z < 0$).
On the other hand, the electrons of the S$^+$ channel show a broad distribution and less pronounced asymmetry than those of the OCS$^+$ channel at $\Delta \phi = \pi$.
The electron momentum distributions vary depending on channel and $\Delta\phi$, suggesting that different electron ejection processes take place in the OCS$^+$ and S$^+$ channels.

\subsection{Asymmetry of electron ejection direction}

To evaluate the asymmetry in the electron ejection direction quantitatively, the asymmetry parameter $\alpha$ is defined as
\begin{equation}
    \alpha(E_\text{kin}, \Delta\phi) = \frac{Y_+(E_\text{kin}, \Delta\phi) - Y_-(E_\text{kin}, \Delta\phi)}{Y_+(E_\text{kin}, \Delta\phi) + Y_-(E_\text{kin}, \Delta\phi)},
\end{equation}
where $Y_+$ and $Y_-$ are yields of electrons with positive and negative momenta along the $z$-axis, respectively.
The yields are obtained as the number of electrons ejected within a polar angle of \SI{30}{\degree} with respect to the laser polarization direction.

\subsubsection{Asymmetry in the OCS$^+$ channel}
\begin{figure}
    \centering
    \includegraphics[width=\hsize]{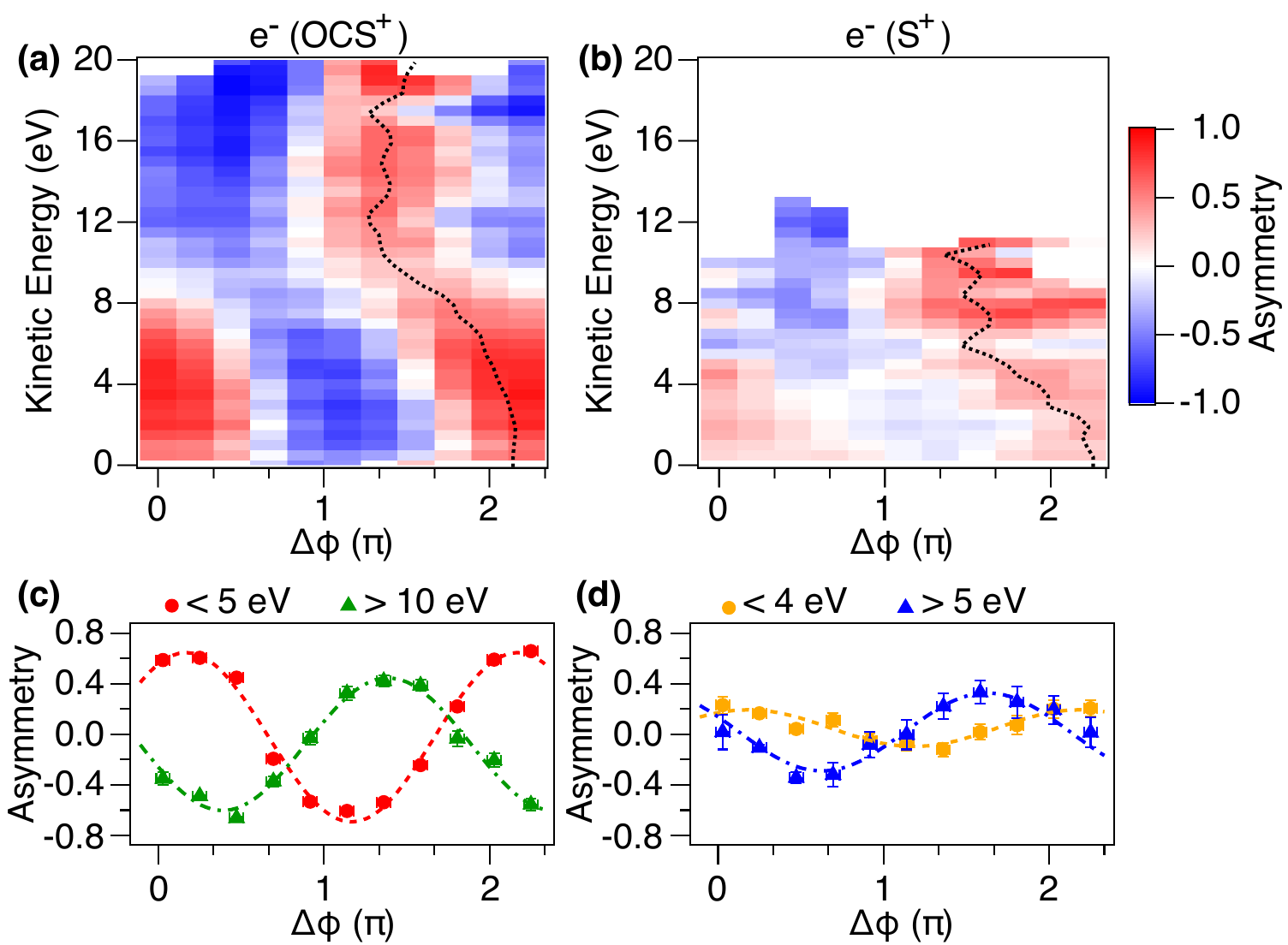}
    \caption{
    Two-dimensional plots of the asymmetry parameter for (a) the OCS$^+$ channel and (b) the S$^+$ channel.
    Dotted lines indicate the phase offset $\eta$ at which the asymmetry parameter has a maximum value at each kinetic energy.
    (c) Integrated asymmetry parameters of the OCS$^+$ channel over the ranges of 0-5 eV (red circles) and 10-20 eV (green triangles). (d) Integrated asymmetry parameters of the S$^+$ channel over the ranges of 0-4 eV (orange circles) and 5-12 eV (blue triangles). Lines represents the fitting results using a cosine function.
    }
    \label{fig:asymmetry}
\end{figure}

Figure~\ref{fig:asymmetry}(a) shows the two-dimensional plots of $\alpha(E_\text{kin}, \Delta\phi)$ of the OCS$^+$ channel.
The asymmetry parameter shows a $2\pi$-oscillatory behavior depending on $\Delta \phi$.
In addition, the asymmetry parameter flips at the kinetic energy of \SI{8}{\eV} at $\Delta \phi = 0$.

The $2\pi$-oscillatory behavior depending on $\Delta\phi$ can be explained by a semi-classical picture of a tunneling electron in alternating electric fields.
Under the assumptions that the initial momentum is zero and the effects of the Coulomb interaction are negligible, the final momentum of the tunneling electron $\mathbf{p}_f$ generated at the ionization time of $t_i$ is obtained by solving Newton's equation of motion,
\begin{equation}
    \mathbf{p}_f(t_i) = -\int_{t_i}^{\infty} F(t, \Delta\phi) dt = \mathbf{A}(t_i, \Delta\phi), \label{eq:vectorpotential}
\end{equation}
where $\mathbf{A}(t_i, \Delta\phi)$ is the vector potential,
which predicts an asymmetric electron momentum distribution along the laser polarization direction depending on $\Delta \phi$.
According to Eq.~(\ref{eq:vectorpotential}), the asymmetry parameter exhibits symmetric electron ejection at $\Delta \phi = 0$ and $\pi$ unlike the experimental results.
This discrepancy can be attributed to the effects of the Coulomb potential discussed in Section 4.2.2.

The asymmetry flip at $E_\text{kin} = \SI{8}{\eV}$ can be attributed to a change in the main contribution from the forward- to backward-scattered electrons.
Since the electrons ejected near the peak of the electric field amplitude recollide to the parent ion at the time when the amplitude is close to 0, the electric field direction flips before and after the collision. 
Thus, after the collision, the forward-scattered electron is decelerated while the backward-scattered electron is accelerated.
As a result of the entire interaction with the laser field, the backward-scattered electron has a higher kinetic energy and is ejected to the opposite direction compared to the forward-scattered electron.
Under the same assumptions of zero initial momentum and no Coulomb interaction, classical trajectory simulations show that the maximum kinetic energy of the forward-scattered electron varies between $2.2U_p$ (6.6~eV) and $2.6U_p$ (7.8~eV) depending on $\Delta\phi$ in the present experimental condition ($I_{2\omega}/I_{\omega} = 0.1$).
Thus, the electrons with the kinetic energy below $2.2U_p$ and above $2.6U_p$ are mainly assigned to the forward- and backward-scattered electrons, respectively, consistent with the observed asymmetry flip around 8~eV.

To clarify the $\Delta \phi$ dependence of two kinetic energy regions, the $E_\text{kin}$-integrated asymmetry parameters are shown in Fig.~\ref{fig:asymmetry}(c).
The electrons in the low-energy region ($E_\text{kin} < \SI{5}{\eV}$) are preferentially ejected to the larger amplitude side of the laser electric fields; the asymmetry parameter shows the positive values at $\Delta \phi = 0$.
On the other hand, the electrons in the high-energy region ($E_\text{kin} > \SI{10}{\eV}$) are preferentially ejected to the lower amplitude side of the laser electric fields; the asymmetry parameter shows the negative values at $\Delta \phi = 0$.

For a quantitative evaluation, the $E_\text{kin}$-integrated asymmetry parameter $\alpha(\Delta \phi)$ is fitted with a cosine function;
\begin{eqnarray}
    A_0 \cos \left(\Delta \phi - \eta \right), \label{eq:eta_fit}
\end{eqnarray}
where $A_0$ is the amplitude of $\alpha(\Delta \phi)$ and $\eta$ is the phase offset, where $\alpha(\Delta \phi = \eta)$ has a maximum value.
The amplitude $A_0$ and the phase offsets $\eta$ are 0.67 and $2.17\pi$ in the low-energy region and 0.52 and $1.39\pi$ in the high-energy region, respectively.
The shift in the phase offsets of the two energy regions is $\Delta\eta = \eta_\text{high} - \eta_\text{low} = -0.78\pi$, which reflects a flip in the asymmetry of the electron emission direction due to the interplay of forward- and backward-scattering, as discussed above in the absence of the Coulomb potential.
The measured phase offset of $-0.78\pi$, instead of the ideal value of $-\pi$, can be attributed to the interaction with the three-center Coulomb potential of OCS$^+$, rather than with a pure one-center Coulomb potential (see Fig.~\ref{fig:CTMC_Coulomb} for the latter).
To determine the asymmetry flipping energy $E_\text{flip}$ at which the dominant scattering process is changed from forward to backward, the two-dimensional asymmetry parameter $\alpha(E_\text{kin}, \Delta \phi)$ is fitted with Eq.~(\ref{eq:eta_fit}) at each electron kinetic energy.
The resulting phase offset $\eta(E_\text{kin})$ is plotted as a dotted line in Fig.~\ref{fig:asymmetry}(a).
The asymmetry flipping energy $E_\text{flip}$ is defined as the kinetic energy where a value of $\eta(E_\text{kin})$ is the middle point between those in the low- and high-energy regions.
That is, $\eta(E_\text{flip}) = 1.78\pi$, and $E_\text{flip}$ is evaluated to be $\SI{8.2}{\eV}$ in the OCS$^+$ channel.

\subsubsection{Effects of Coulomb potential on the asymmetry}
Here, the effects of Coulomb potential on the asymmetry parameter are briefly described with a pure one-center Coulomb potential.

\begin{figure}
    \centering
    \includegraphics[width=\hsize]{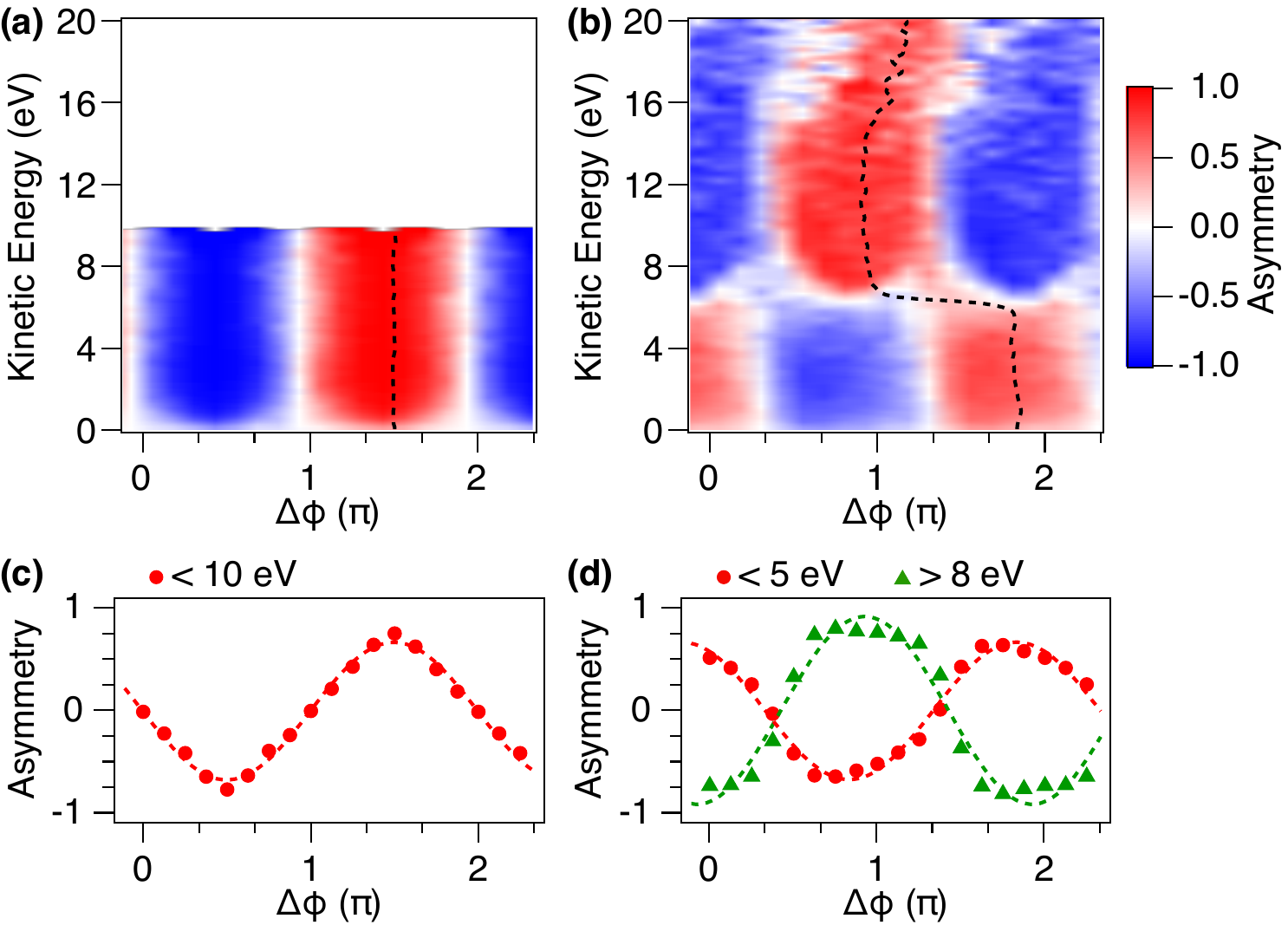}
    \caption{
    Simulated two-dimensional plots of the asymmetry parameter (a) without any Coulomb potential and (b) with a pure one-center Coulomb potential in phase-locked $\omega + 2\omega$ laser fields ($I_\omega = \SI{3.5e13}{\Wcm}$, $I_{2\omega} = \SI{3.5e12}{\Wcm}$).
    (c) Integrated asymmetry parameters simulated without any Coulomb potential over the ranges of 0-10 eV (red circles). (d) Integrated asymmetry parameters simulated with a pure one-center Coulomb potential over the ranges of 0-5 eV (red circles) and 8-15 eV (green triangles).
    }
    \label{fig:CTMC_Coulomb}
\end{figure}

The two-dimensional plot of the simulated asymmetry parameter in the phase-locked $\omega + 2\omega$ laser fields ($I_{\omega} = \SI{3.5e13}{\Wcm}, I_{2\omega} = \SI{3.5e12}{\Wcm}$) without any Coulomb potential is shown in Fig.~\ref{fig:CTMC_Coulomb}(a).
As described by Eq.~(\ref{eq:vectorpotential}), the asymmetry parameter has a maximum value at $\Delta \phi = 1.5\pi$ and the electrons are ejected symmetrically at $\Delta \phi = 0, \pi$, and $2\pi$ as shown in Fig.~\ref{fig:CTMC_Coulomb}(c).
In addition, since electron scattering by a Coulomb potential is not taken into account, the asymmetry parameter is independent of the kinetic energy of the electron.

The two-dimensional plot of the asymmetry parameter simulated with a pure one-center Coulomb potential $V_\text{pure}$ is shown in Fig.~\ref{fig:CTMC_Coulomb}(b).
The pure one-center Coulomb potential is given by
\begin{equation}
    V_\text{pure} = -\frac{1}{\sqrt{r^2 + z^2}}.
\end{equation}
When the Coulomb potential is taken into account, the two-dimensional plot is clearly different from that simulated without the Coulomb potential.
The asymmetry parameter shows a clear electron kinetic energy dependence, in which the electrons in the low- and high-energy regions are attributed to the forward- and backward-scattered electrons by the Coulomb potential~\cite{Hao.PRA.2020}.
In addition, the asymmetry parameter has a maximum value at $\Delta \phi = 1.82\pi$ and $0.92\pi$ in the low-energy ($E_\text{kin} < \SI{5}{\eV}$) and high-energy ($E_\text{kin} > \SI{8}{\eV}$) regions as shown in Fig.~\ref{fig:CTMC_Coulomb}(d), respectively, instead of $1.5\pi$.

These results indicate that the electron scattering process by the Coulomb potential causes the dependence on the electron kinetic energy and the shift of $\Delta \phi$ dependence of the asymmetry parameter.
There are a significant discrepancy between the experimental results and the simulated results with the pure one-center Coulomb potential.
To evaluate the effect of the Coulomb potential quantitatively, we perform the numerical CTMC simulation including a DFT potential in Sec. 4.3.

\subsubsection{Asymmetry in the S$^+$ channel}
The two-dimensional plot of the asymmetry parameter of the S$^+$ channel is shown in Fig.~\ref{fig:asymmetry}(b), indicating a clear difference from that of the OCS$^+$ channel.
In the S$^+$ channel, the asymmetry is weaker across the entire energy range and flips at lower kinetic energy than that in the OCS$^+$ channel.
The larger asymmetry is observed in a high-energy region than in a low-energy region, in contrast to the OCS$^+$ channel.
The amplitude $A_0$ and the phase offsets $\eta$ are 0.14 and $2.18\pi$ in the low-energy ($E_\text{kin} < \SI{4}{\eV}$) region and 0.31 and $1.63\pi$ in the high-energy ($E_\text{kin} > \SI{5}{\eV}$) region, respectively, as shown in Fig.~\ref{fig:asymmetry}(d).
The shift in the phase offsets between the low- and high-energy regions is $\Delta\eta = -0.55\pi$, which is $0.23\pi$ larger than $\Delta\eta = -0.78\pi$ in the OCS$^+$ channel.

Significant drops of the observed asymmetry in the low-energy region and larger deviation of $\Delta \eta$ from $-\pi$ in the S$^+$ channel than in the OCS$^+$ channel suggest that the mean electron trajectories deviate from those expected with a pure one-center Coulomb potential.
This is consistent with a rescattering excitation picture, in which electron trajectories in the OCS$^+$ channel, including non-scattered and elastically scattered electrons, tend to pass farther from the ion core, whereas those in the S$^+$ channel, involving inelastically scattered electrons, pass closer to the core.
The Coulomb potential of OCS$^+$ approaches the $1/r$ potential at large distances from the ion core, but deviates significantly at small distances due to its three-center structure.
As a result, the influence of the three-center Coulomb potential becomes more pronounced, leading to stronger distortion of electron trajectories compared with those expected for a pure one-center Coulomb potential in the S$^+$ channel.

The asymmetry flipping energy of $E_\text{flip} = \SI{4.2}{\eV}$ in the S$^+$ channel is \SI{4}{\eV} lower than that in the OCS$^+$ channel, which is consistent with the fact that the cross section for the photodissociation of OCS$^+$ has a maximum at approximately 4.0 eV~\cite{Orth.CP.1980}.
This shift of $E_\text{flip}$ suggests that the molecular excitation occurs by the energy transfer in inelastic scattering of the ejected electron to the ion.

Other mechanisms than the mechanism III, tunneling ionization and electron recollisional excitation, are not expected to contribute the significant drops of the asymmetry in the electron momentum distribution and the shift of $E_\text{flip}$ of the S$^+$ channel.
For the mechanism I, multi-photon excitation following photoelectron emission, the electron emission is not affected by the excitation process, which would lead to similar behaviors of the electron momentum distribution in both OCS$^+$ and S$^+$ channels.
For the mechanisms II, tunneling ionization from HOMO-1, neither the ionization potential nor the shape of molecular orbital significantly affects the electron asymmetry, as the electron trajectories are primarily governed by the shape of the applied laser electric field.
The $E_\text{flip}$ corresponds to the maximum kinetic energy of the forward-scattered electrons and is independent of the orbital from which an electron is ejected, such as HOMO and HOMO-1.
In addition, the tunneling ionization rate from HOMO-1 is estimated to be more than three orders of magnitude slower than that from HOMO in the present experimental conditions.

From the above discussions, the mechanism III is considered to be most likely to occur.
For more detailed evaluation, we perform the CTMC simulation.

\begin{figure*}
    \centering
    \includegraphics[width=0.7\hsize]{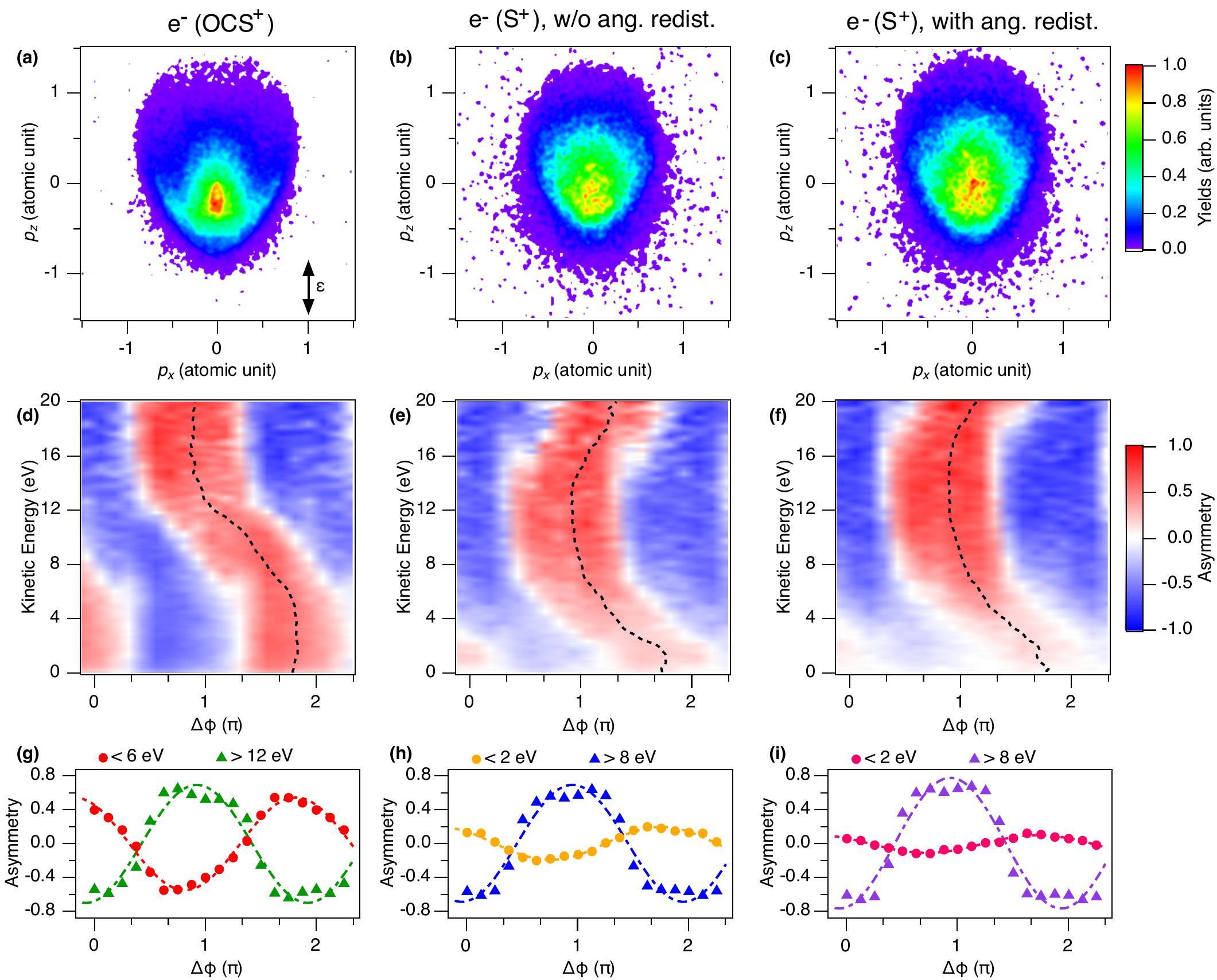}
    \caption{Results of the CTMC simulation.
    (a) Simulated momentum image of the photoelectron corresponding to the OCS$^+$ channel in phase-locked $\omega + 2\omega$ laser fields ($I_{\omega} = \SI{3.5e13}{\Wcm}, I_{2\omega} = \SI{3.5e12}{\Wcm}$) at $\Delta\phi = \pi$. The laser polarization direction is indicated as $\varepsilon$. (b), (c) Same as (a), but corresponding to the S$^+$ channel without and with angular redistribution due to inelastic scattering. 
    (d) Simulated two-dimensional plots of the asymmetry parameter of the OCS$^+$ channel.
    Dotted lines indicate the phase offset $\eta$ at which the asymmetry parameter has a maximum value at each kinetic energy.
    (e), (f) Same as (d), but for the S$^+$ channel without and with angular redistribution.
    (g) Integrated asymmetry parameters of the OCS$^+$ channel over the ranges of 0-6 eV (red circles) and 12-20 eV (green triangles).
    (h), (i) Same as (g), but for the S$^+$ channel without and with angular redistribution over the ranges of 0-2 eV (orange and magenta circles) and 8-16 eV (blue and purple triangles). Lines represents the fitting results using a cosine function.
    }
    \label{fig:CTMC}
\end{figure*}

\subsection{CTMC simulation}
The simulated results of electron trajectories in the phase-locked $\omega + 2\omega$ laser fields ($I_{\omega} = \SI{3.5e13}{\Wcm}, I_{2\omega} = \SI{3.5e12}{\Wcm}$) are summarized in Fig.~\ref{fig:CTMC}.

\subsubsection{Simulated momentum images}
The simulated momentum image corresponding to the OCS$^+$ channel at $\Delta\phi = \pi$ is shown in Fig.~\ref{fig:CTMC}(a).
A strong asymmetric distribution along the laser polarization direction $\varepsilon$ is obtained as clearly as that in Fig.~\ref{fig:momimage}(c).
The electrons are preferentially ejected to the larger amplitude side of the $\omega + 2\omega$ laser fields ($p_z < 0$).
No periodic peak structure is observed because the classical trajectory simulation is employed and the interference between electron wavepackets emerging every optical cycle is not included.
Reproduction of the series of peak structures is beyond the scope of this study.

The simulated momentum image corresponding to the S$^+$ channel, where the inelastic scattering between the electron and the ion is included, is shown in Fig.~\ref{fig:CTMC}(b).
The electrons show a weak asymmetric distribution along the laser polarization direction in contrast to the OCS$^+$ channel.
Fig.~\ref{fig:CTMC}(c) includes angular redistribution by inelastic scattering, whereas Fig.~\ref{fig:CTMC}(b) does not.
Both momentum images exhibit similar distributions.
Only a slight expansion of the momentum distribution is recognized in the random scattering model, which means that electrons have slightly higher kinetic energy due to angular redistribution.
Irrespective of the angular redistribution, the CTMC momentum images qualitatively reproduce the experimental results that the electron is preferentially ejected to the larger amplitude side in the OCS$^+$ channel and weak asymmetry is observed in the S$^+$ channel.

\subsubsection{Simulated asymmetry parameters}
The two-dimensional plots of the simulated asymmetry parameter $\alpha(E_\text{kin}, \Delta\phi)$ of both channels are shown in Figs.~\ref{fig:CTMC}(d), (e) and (f).
As observed in the experiment, the simulated $\alpha(E_\text{kin}, \Delta\phi)$ shows clear dependencies on $\Delta \phi$ and $E_\text{kin}$; $2\pi$-oscillatory behavior and asymmetry flip.
To evaluate the phase offset $\eta$ in the low- and high-energy regions, the $E_\text{kin}$-integrated asymmetry parameters $\alpha(\Delta \phi)$ are shown in Figs.~\ref{fig:CTMC}(g), (h) and (i).

In the OCS$^+$ channel, the asymmetry amplitudes $A_0$ and the phase offsets $\eta$ are 0.55 and 1.81$\pi$ in the low-energy region ($E_\text{kin} < \SI{6}{\eV}$), and changing to 0.70 and 0.94$\pi$ in the high-energy region ($E_\text{kin} > \SI{12}{\eV}$).
Although absolute values of the phase offsets $\eta$ in the simulation are lower than those observed in the experiment by more than $0.3\pi$, the shift in the phase offsets of $\Delta\eta = -0.88\pi$ agrees more closely with the experimental results ($-0.78\pi$) in Fig.~\ref{fig:asymmetry}(c).

In the S$^+$ channel without angular redistribution, the asymmetry amplitudes $A_0$ decrease to 0.19 in the low-energy region ($E_\text{kin} < \SI{2}{\eV}$) and remain at 0.69 in the high-energy region ($E_\text{kin} > \SI{8}{\eV}$).
This significant drop of the asymmetry amplitude in the low-energy region is consistent with the experimental observations (Fig.~\ref{fig:asymmetry}).
The phase offsets $\eta$ are $1.74\pi$ and $0.95\pi$ for the low- and high-energy regions, respectively.
When angular redistribution is employed, the amplitudes further decrease to 0.10 in the low-energy region but increase to 0.77 in the high-energy region, whereas the phase offsets $\eta$ ($1.73\pi$ and $0.94\pi$) remain unchanged.
This indicates that the angular redistribution by inelastic scattering plays a minor role, whereas the electron trajectories in intense laser fields are governed primarily by the interaction with the laser electric field and the Coulomb field throughout the entire trajectory.
The shift of the phase offsets $\Delta\eta = -0.79\pi$ in the S$^+$ channel is smaller compared to the OCS$^+$ channel.
This trend is also consistent with the observed results in the experiments.

The asymmetry flipping energy $E_{\rm flip}$ is \SI{10.2}{\eV} in the OCS$^+$ channel, but decreases to 3.7 eV and 4.3 eV in the S$^+$ channel without and with angular redistribution, respectively.
Upon closer inspection, the simulated $\eta(E_\text{kin})$ in the OCS$^+$ and S$^+$ channels (black dotted lines in Figs.~\ref{fig:CTMC}(d), (e) and (f)) exhibits a more gradual variation with electron kinetic energy, whereas the experimental results show a more rapid variation.
As shown in Fig.~\ref{fig:CTMC_Coulomb}(b), the CTMC simulation with a pure Coulomb potential exhibits a step-like behavior.
The experimental results fall between the two CTMC simulations, which may suggest that the effective potential differs from the static DFT-Coulomb combined potential.

Overall, the CTMC simulations qualitatively reproduce the main experimental features, including the direction of electron ejection, the reduction of asymmetry in the S$^+$ channel, and the relative shift of $E_\text{flip}$ between the OCS$^+$ and S$^+$ channels, indicating that the tunneling ionization and recollisional excitation play important roles in molecular dissociation processes in the intense laser fields.
However, there are quantitative discrepancies between the experimental and simulated results, such as absolute values of $\eta$, $E_\text{flip}$, and $\Delta \eta$ in the OCS$^+$ and S$^+$ channels.
These discrepancies suggest that further improvement of the potential energy surfaces is needed, including refining the approach to combining the DFT and Coulomb potentials and accounting for electron excitation and redistribution induced by inelastic scattering.
Electron recollisional excitation modifies the electronic wave function of OCS$^+$, transforming the potential energy surface into that of the excited state.
In addition, in the present CTMC simulation, the volume effects are not considered, which affect not only the effective laser field intensity and its spatial distribution but also $\Delta \phi$ due to the Gouy phase.

In laser fields with higher intensity, the electron trajectories are predominantly governed by the laser electric field, and the influence of the Coulomb potential becomes weaker.
As a result, the dependence on the electron kinetic energy $E_\text{kin}$ of the asymmetry parameter $\Delta \eta$ is expected to be reduced at higher intensities.
Considering that fragment ions are more likely to be generated in a high intensity region because of increase in the recolliding energy of the tunneling electrons, the difference in intensity would be responsible for the larger $\Delta \eta$ in the S$^+$ channel than in the OCS$^+$ channel.

The Gouy phase shift can be written as
\begin{equation}
    \phi_\text{Gouy} = - \arctan \left( \frac{x}{x_R} \right),
\end{equation}
where $x_R= \pi w_0^2/\lambda$ is the Rayleigh length with the beam waist $w_0$, and the wavelength $\lambda$.
The laser pulses are focused at $x = 0$ and the Rayleigh lengths are estimated to be about 2.5 mm for the fundamental pulse and 1.3 mm for the second harmonic pulse, which are comparable with the size of molecular beam of 2 mm.
Thus, the phase difference $\Delta \phi$ between the $\omega$ and $2\omega$ pulses depends on the $x$-position along the laser propagation direction.
The difference in $\phi_\text{Gouy}$ between the $\omega$ and $2\omega$ pulses is about $0.1\pi$ at the largest.
The Gouy phase can also worsen the accuracy of the experimental phase difference beyond the phase stability limit of $0.06\pi$, contributing to a larger discrepancy in $\eta$ between the experiment and simulation.

Other concerns are about the assumptions in the CTMC simulation, such as assuming the initial position and momentum ($r(t_i) = 0, v_z(t_i) = 0$), neglecting a motion of nuclei, and combining the DFT and Coulomb potentials of OCS$^+$.
Especially, the shape of the Coulomb potential in the simulation causes discrepancies in the absolute values of the experimental and simulated phase offset $\eta$ and asymmetry flipping energy $E_\text{flip}$ (see Fig.~\ref{fig:CTMC_Coulomb}).
Nevertheless our simulation using a simple model reproduces the experimental results qualitatively well.

\section*{Conclusions}
In this study, we have investigated electron recollisional excitation leading to molecular dissociation of OCS$^+$ in the phase-locked $\omega + 2\omega$ intense laser fields.
Three-dimensional momenta of photoelectrons show clear asymmetric distribution depending on the phase difference $\Delta\phi$ of the two electric fields, and the photoelectron ejection direction is flipped at a specific photoelectron energy $E_\text{flip}$.
The electrons in the lower and higher kinetic energy than $E_\text{flip}$ are attributed to the forward- and backward-scattered electrons, respectively, showing that the forward-scattered electrons are preferentially emitted toward the larger amplitude side of the laser electric field and the backward-scattered electrons are emitted to the opposite side through the inversion of the photoelectron momentum direction at recollision.
In addition, the asymmetry of photoelectrons also depends on the coincidentally produced cations, OCS$^+$ and S$^+$.
The asymmetry flipping energy $E_\text{flip}$ of the OCS$^+$ channel and the S$^+$ channel were measured to be \SI{8.2}{\eV} and \SI{4.2}{\eV}, respectively, whose difference of \SI{4}{eV} is consistent with the excitation energy of OCS$^+$ from the $X^2\Pi$ to the $A^2\Pi$ states.
The CTMC simulation with the DFT potential has been performed and describes the experimental results well, including the difference in the asymmetry flipping energy $E_\text{flip}$ and the phase offset $\eta$ of both channels.

These results indicate that the molecules are excited by electron recollision in the phase-locked $\omega + 2\omega$ laser fields and that the excited states generated in the intense laser fields could be clarified on the basis of the observed asymmetry flipping energy $E_\text{flip}$.
Our approach by a combination of the PEPICO momentum imaging with the phase-locked $\omega + 2\omega$ laser fields provides a clue of the excited state assignment and enables us to discuss molecular dissociation dynamics induced by inelastic elerctron scattering in intense laser fields.


\section*{Author contributions}
T. E.: Conceptualization, Investigation, Formal analysis, Writing - original draft.
T. O.: Formal analysis, Writing - review \& editing.
R. I.: Supervision, Writing - review \& editing.
All authors approved the final version of the manuscript.

\section*{Conflicts of interest}
There are no conflicts to declare.

\section*{Data availability}
The data supporting the findings of this study are available within the article. Additional raw data are available from the corresponding author upon reasonable request.

\section*{Acknowledgements}
This work was partly supported by JSPS KAKENHI Grant Numbers 19K15515, 22K14654, 22H02043, 23K23311.
The authors are grateful to T. Morishita (UEC) for valuable discussions, Y. Hagihara for technical support, and J. Cao for data acquisition.



\balance


\bibliography{OCS_PEPICO} 
\bibliographystyle{rsc} 

\end{document}